\documentstyle[preprint,prl,aps]{revtex}
\def\be{\begin{equation}}
\def\ee{\end{equation}}
\def\bea{\begin{eqnarray}}
\def\eea{\end{eqnarray}}
\begin{document}
\draft
%@twocolumnfalse\endcsname]
\title{Sterile Neutrino and Accelerating Universe}
\author{P. Q. Hung\cite{email}}
\address{Dept. of Physics, University of Virginia, \\
382 McCormick Road, P. O. Box 400714, Charlottesville, Virginia 22904-4714}
\date{\today}
\maketitle
\begin{abstract}
If all three neutrino oscillation data were to be confirmed in the
near future, it is probable that one might need a sterile neutrino,
in addition to the three active ones. This sterile neutrino, $\nu_S$,
would be very light with mass $m_{\nu_S} \leq 1 eV$ or even
with $m_{\nu_S} \sim 10^{-3} eV$ according to some scenarios. Why
would it be so light?
On another front, recent cosmological observations and analyses
appear to indicate that the
present universe is flat and accelerating, and 
that the present energy density is dominated by a ``dark variety'',
with $\rho_V \sim (10^{-3}eV)^4$. Is it a constant? Is there a link 
between these apparently unrelated phenomena? 

\end{abstract}

%\narrowtext

For the past three years or so, a number of discoveries 
in particle physics and
cosmology has begun to reveal startling results which, 
if verified, would have deep implications. 
Of particular relevance to this paper are
the new results on neutrino oscillation which 
suggest that neutrinos have a
mass, albeit a tiny one, and new evidence for an accelerating universe. 

On the cosmology front, there is a most recent evidence for a flat
universe, i.e. $\Omega = 1$, from the Boomerang collaboration \cite{boom}. 
In addition, discoveries of high red-shift (high Z) Supernovae IA
and their use in determining the deceleration parameter $q_0$ have been
most dramatic \cite{super}. A {\em positive} $q_0$ 
implies that the universe is
decelerating while a {\em negative} one- as implied
by the high Z Supernovae data- means that it is {\em
accelerating}. Furthermore,
assuming that $\Omega_0= \Omega_{m} + \Omega_{\Lambda} =1$, 
where $\Omega_{m, \Lambda}$ refer to the $\Omega$'s coming from 
matter and a cosmological constant respectively,
Supernovae (SNIA) results seem to indicate that $\Omega_{m} \sim 0.3$
and $\Omega_{\Lambda} \sim 0.7$, which although differ from one
another by a factor of two,
are of the same order of magnitude. 
The vacuum energy density coming from a cosmological constant
would be approximately $\rho_{V} \sim (1.6 \times 10^{-3} eV)^4$. 
Why is it so small?

There are several
appealing suggestions for such a ``tiny'' 
(although presently dominant) value for the vacuum energy
(which come with their own difficulties), among which
is the idea of a dynamical vacuum energy, or in other words ``quintessence''
\cite{quint}.
If, on the other hand, this vacuum energy were to genuinely come from 
some phase transition, there can be 
interesting consequences, such as the links with current physical
phenomena as presented in this paper. With
$\rho_V = V(0) = \sigma^4$, one
would then expect $\sigma \sim 1.6 \times 10^{-3} eV$. 
%There are a couple of interesting possibilities which are
%worth investigating. For example, one can 
%either assume that $A^{1/4} \sim O(1)$ and $v \sim 10^{-3} eV$,
%or $A^{1/4} < 1$ and $v > 10^{-3} eV$. The extreme fine tuning
%case is when $A^{1/4} \ll 1$ and $v \gg 10^{-3} eV$. These 
%different choices will of course have very different physical
What might be the origin of  of such a small $\sigma$? 
In the Standard Model,
there are several symmetry breaking scales, e.g. the electroweak and the
chiral symmetry breaking scales, each of which contributes a cosmological
constant several orders of magnitude larger than the aforementioned
constant. One generally agrees that a cancellation of something like
$\rho_{v} \sim 4 \times 10^{45} eV^4$ (Electroweak)
down to $\rho_{v} \sim 10^{-12} eV^4$ is highly unnatural. 
In the present {\em absence of a satisfactory solution} to
the cosmological constant problem,
one might then assume that there is a yet-unknown mechanism
by which the various cosmological
constants get cancelled out to zero, and the hint that one might
have a non-zero vacuum energy at the present time indicates some
new contribution, either in the form of quintessence 
or a genuinely ongoing
new phase transition. It is the latter assumption that we shall
exploit in this paper. In other words, this unknown mechanism would
bring the total vacuum energy down to zero when {\em all} phase
transitions are completed. 

%What does neutrino oscillation have to do with an accelerated universe? At first
%thought, it appears to be ridiculous to even think that these two phenomena have
%anything to do with one another. Upon close examination of all three hints
%of a neutrino mass (solar, atmostpheric and LSND results), there appears to be
%a link between these two seemingly disparate phenomena. Let us briefly describe 
%this point before going into more details. 

On another front, it is well accepted that if 
all three neutrino oscillation results 
(solar, atmostpherics, and LSND data) were to be proven correct, 
one would need, in addition to
three standard neutrinos, $\nu_e$, $\nu_{\mu}$, and $\nu_{\tau}$,
an additional one which does not have normal
electroweak interactions- the so-called sterile neutrino $\nu_{S}$
\cite{neutrino}. This neutrino would have a mass 
$m_{\nu_S} \leq 1 eV$ or even $m_{\nu_S} \sim 10^{-3} eV$ 
according to some scenarios. In order to explain the data, this
neutrino would mix with one or more active neutrinos.
%One of 
%the scenarios which fits all three oscillation results goes approximately 
%like this. 
%For example, a scenario with four neutrinos (three active and one sterile)
%A scenario which fits all three oscillation data goes as follows.
%The mass eigenvalues of the neutrino mass matrix
%have the following pattern ($\nu_{0,1,2,3}$ being mass eigenstates):
%$\nu_2, \nu_3 \approx (1/\sqrt{2})(\nu_{\mu} \pm \nu_{\tau})$ 
%have a mass of approximately $1 eV$ with
%a mass splitting $\Delta m^2$ of $O(10^{-3} eV^2)$ (atmospheric data), 
%and $\nu_1 \approx \nu_{e}$ and $\nu_0 \approx \nu_{S}$ have a mass 
%of approximately $10^{-3} eV$, 
%with $\Delta m^2$ of $O(10^{-5} eV^2)$ (solar MSW), with $\Delta m^2$ 
%between the ``large' and the ``small'' eigenvalues being of 
%$O(1 eV^2)$ (LSND). 
%Much work will be 
%needed to verify if all three oscillation results are indeed correct (especially
%the LSND result), and as a result whether or not one would need a 
%sterile neutrino. 
If the idea of a sterile neutrino proves to be correct in the future, 
one will
be confronted with the following puzzling question: Why is
$\nu_S$ so light and so close in mass to an active neutrino
when it appears that they are of
very different types of particles? 
In popular scenarios with the see-saw mechanism and
Majorana mass, typically the sterile neutrino is {\em heavy} 
and the scale of new
physics is rather large (a typical Grand Unified scale).
Needless to say, the issues of neutrino mass are far from being 
resolved, including the important question of whether or not 
the mass is Dirac or Majorana. One should finally also
notice that there are additional
(astrophysical) arguments which claim the need for a
sterile neutrino \cite{astro}.
%Even if the suggestions that
%the electron neutrino oscillates into a sterile neutrino in
%order to solve the solar neutrino problem proves to be 
%incorrect- one might not need a sterile neutrino if the LSND
%results are not confirmed, there appears to be other 
%astrophysical arguments suggesting
%the need for its existence. 

What we would like to propose in this article is a ``simple''
model which links the issue of the origin of the 
sterile neutrino mass to
that of the dark energy. In a nutshell, it is simply
this: the sterile neutrino obtains its mass through a Yukawa
coupling with a ``singlet'' Higgs field whose effective potential
is of a ``slow-rolling'' type. During this slow rolling, 
the vacuum energy would 
be given approximately by $V(0) = \sigma^4$. In our model,
the {\em effective} mass of the
sterile neutrino is proportional to the {\em present} value
of the singlet Higgs field, $\phi_S (t_0)$ and can be as small
as $10^{-3} eV$ provided $\phi_S (t_0)$ is itself sufficiently small. 
The Higgs field $\phi_S$ will
eventually reach its global value $v_S$ which can 
be {\em much greater} than $\phi_S (t_0)$. 
In this sense, a small sterile neutrino mass at the present
time is, in our scenario, merely a reflection of the current
value of $\phi_S$. At the end of the phase transition, the
sterile neutrino could, in principle, acquire a 
{\em much larger mass}. It is an intriguing possibility
that, as $\phi_S$ evolves, the sterile neutrino mass will
change and, as a result, the oscillation with the electron
neutrino will end in some distant future.

%Nevertheless, let us imagine for the moment that 
%one does need a sterile neutrino and that its
%mass is of $O(10^{-3}-10^{-2} eV)$. 
%What might be the origin of such a tiny mass for the sterile neutrino? 
%Being ``sterile'', this neutrino most likely obtains its mass from a
%source which is {\em very different} from the other three ``active'' 
%neutrinos. The scale of physics which could give rise to the sterile
%neutrino mass might even be {\em tiny}.
%Is it possible that the origin of this tiny mass might be linked to
%the present vacuum energy?
%There exists many models
%for light neutrino masses which can be either Majorana or pur Dirac in
%nature. (The question whether
%or not the light neutrino mass is Majorana or Dirac is an extremely important 
%one.) However, in all of those models, it turns out that it is extremely hard
%(at least in the present context) to construct a scenario in which the sterile
%neutrino is so light. 
%It is
%for this reason that many model builders prefer a wait-and-see attitude towards
%the LSND result and are usually content with trying to explain mainly the
%solar and atmospheric neutrino oscillation results. 

%So, what if the idea of a sterile 
%neutrino turns out to be tenable (e.g. if all three oscillation results
%were proved to be correct)? What will the scale of the new physics which 
%would give rise to such a tiny mass be? 

In what follows, the active neutrinos will be assumed
to obtain their masses by phase transitions which have
{\em already} occured. These masses could either be
Majorana or Dirac masses. (One example of a naturally light
Dirac neutrino mass can be found in \cite{hung}, along with
numerous phenomenological consequences.) The singlet sterile
neutrino, on the other hand, will be assumed to obtain 
primarily a mass
through a Yukawa coupling with a singlet Higgs field as 
described above. In other words,
this mass would be obtained during the ``last''
phase transition.
%Here, we will also assume that the
%sterile neutrino has a Dirac mass.

%The paper will follow the 
%same trend of logic used in the construction of a model of 
%small Dirac neutrino
%masses \cite{hung}. There, it was found that one can naturally obtain tiny
%Dirac neutrino masses in a model with numerous phenomenological
%consequences. Here, we will assume that the sterile
%neutrino also has a Dirac mass. 
%This assumption turns out to
%fit quite conveniently with our scenario.
%To simplify the discussion,
%we shall first ignore any possible mixing between $\nu_S$ and, say, $\nu_e$
%(which would be needed in order to explain the solar oscillation data).
%This will be dealt with at the end of the manuscript.

First, a phenomenological effective potential for 
a singlet Higgs field is presented along with a relationship
between the vacuum expectation value, $v_S$, of the singlet Higgs
field and its present classical value, $\phi_S(t_0)$. Second,
we shall use this relationship {\em in conjunction} with the sterile
neutrino mass to infer on a possible magnitude for $v_S$. 

The scenario that we would like to discuss here is the following.
1) There are several phase transitions occuring during the
course of history of the universe. 2) The ``last'' (perhaps)
of those phase transitions is the one associated with $\phi_S$.
3) We will assume that there exists a mechanism by which the vacuum energy
vanishes after the associated phase transition is completed.
As a result, the total energy density is given by
$\rho_{tot} = \rho_m + \rho_V$,
where $\rho_m$ includes matter energy density of all types,
and- this is the crucial assumption- $\rho_V$ is the vacuum energy
which will include, at any given time, the total cosmological constant
arising from phase transitions which either have not started
or have not been completed. 
%This would mean that $\rho_V$ is
%not a fixed number but varies with each stage of transition.
In our scenario, before the ``last''phase transition is completed, 
$\rho_V$ will simply be given by $V(\phi_{S}=0)$.

The {\em phenomenological} effective potential used to illustrate our 
scenario is as follows:
\begin{equation}
V(\phi_S) = \sigma^4 (1- a x^2 + b x^2 exp(-c x^2) + d x^3),
\label{potential}
\end{equation}
where $x \equiv \phi_S^2/v_S^2$ with $v_S$ being the value at the 
global minimum, and where $V(\phi_S=0)= \sigma^4$.
The arbitrary coefficients $a,b,c,d$ are chosen
so as to make $V(\phi_S)$ {\em very flat}. 
Since this potential is purely {\em phenomenological}, issues
such as quantum corrections, etc.., are already included in
the choice of these parameters, i.e. $V(\phi_S)$ is presumably
the ``final'' form arising from some unknown deeper theory.
This peculiar form for the potential is inspired, in parts,
by previous studies of inflationary models \cite{kolb}. 
This potential obeys:
$V(\phi_S = v_S)= 0$, $V^{\prime}(\phi_S = v_S)=0$. For 
a given set of values for $a,c,d$, these conditions restrict $b$ to
be $b = (d/2 -1) exp(c) /c$. 
Let us furthermore assume that, whatever barrier that existed, the
singlet field has already proceeded to {\em classically} ``roll
down'' to its global minimum at $v_S$. 
The equation describing the evolution of $\phi_S(t)$ is
a well-known one, namely
$\ddot{\phi} + 3 H \dot{\phi} + V^{\prime} (\phi_S) = 0$.

One remark about the potential\ (\ref{potential}) is in order 
here. The analysis presented below depends primarily
on two points: 1) $V(0) = \sigma^4$; 2) With
$V(\phi_S) = \sigma^4 \tilde{V}(x)$, the present
slow-rolling requirement would be satisfied if
$2 \tilde{V}(x)^{\prime} + 4 x \tilde{V}(x)^{\prime\prime}
\sim O(x)$. Although we use a potential of the form\
(\ref{potential}), any other potential which satisfies
those two points might also work. One might assume that
there is a class of models where the above conditions
are satisfied. 

The evolution of the scale factor
$R(t)$, under the assumption of zero curvature ($k=0$)
as is presumably the case experimentally, is given by
\begin{equation}
\label{hubble}
H(t)^2 = (8 \pi/3 m_{Pl}^2) (\rho_m + \rho_V),
\end{equation}
where $H(t) \equiv \dot{R} / R$ and
where $m_{Pl}$ is the Planck mass. One can also rewite
the above equation using the definition:
$\Omega(t) \equiv (8 \pi/3 m_{Pl}^2)(\rho/H(t)^2)$,
namely $\Omega_m(t) + \Omega_{\Lambda}(t)= 1$.
From Eq.\ (\ref{hubble}), one observes that,
because $\rho_m$ {\em decreases} with time, the Hubble
parameter $H(t)$ also {\em decreases} and tend towards
$H_{V}^2 \sim (8 \pi/3 m_{Pl}^2) \rho_V$, although
the present value for $H(t_0) \equiv H_0$ does
not differ by much from $H_V$. In fact, with
$\Omega_m \sim 0.3$ and $\Omega_{\Lambda} \sim 0.7$,
it is easy to see that $H_{0}^2 = H_{V}^2 (1 + 
\rho_{m}(t_0)/\rho_V) \sim 1.43 H_{V}^2$. Furthermore,
since $\rho_V$ remains constant, 
$\Omega_{\Lambda}(t)$ {\em increases} with time, implying
that $\Omega_m(t)$
decreases with time. The universe will become more
and more dominated by the vacuum energy. 

From Eq.\ (\ref{hubble}), we can set a constraint on
the parameter $\sigma$ appearing in $V(\phi_S)$.
First, in our scenario, $\rho_V = V(\phi_S =0) =
\sigma^4$.
Denoting the present Hubble rate $H(t_0)$ by $H_0$, 
we can write down the following inequality:
\begin{equation}
\sigma^4 \leq (3 m_{Pl}^2/8 \pi) H_0^2.
\end{equation}
Putting in the present value for $H_0$, one obtains
$\sigma \leq 10^{-3} eV$. 

%The next question concerns an estimate of what the
%value of $v_S$ might be or, more conservatively, what
%its range of values might be. 
What range of values might one expect for $v_S$?
To find this out,
we assume that the present universe is in a stage where
$\phi_S$ is ``slowly rolling''- and this is what the above
form for $V(\phi_S)$ is supposed to do. As stated, it is
irrelevant at this stage to try to determine the exact 
value for $v_S$. Consequently, we will approximate
$H$ to be a constant, namely $H \sim H_V$, noting
that $H_0 \sim 1.2 H_V$ as shown above.
One obtains the usual constraint:
$|V^{\prime\prime}(\phi_{S})| \lesssim 9 H_{V}^2$.
If we write $V(\phi_S) = \sigma^4 \tilde{V}(x)$
with $x \equiv (\phi_S/v_S)^2$, the previous constraint
translates into:
\begin{equation}
\label{Vpp2}
|2 \tilde{V}^{\prime} + 4 x \tilde{V}^{\prime\prime}|
\lesssim 24 \pi (\frac{v_S}{m_{Pl}})^2,
\end{equation}
where the primes denote derivatives with respect to $x$.
If the universe is in the stage where $x \ll 1$, the 
constraint \ (\ref{Vpp2}) can be translated into a
constraint on $v_S$ as a function of what one might
think the present value of $\phi_S$ would be, namely
\begin{equation}
\label{vS}
v_S \gtrsim (\frac{a-b}{2 \pi})^{1/4} \sqrt{\phi_{S,0}
m_{Pl}},
\end{equation}
where $\phi_{S,0}$ refers to the present value of $\phi_S$.
(Notice that this bound is independent of the value of
$\sigma$.)
To be able to make use of the bound \ (\ref{vS}), one
should specify what the parameters $a,b$ are as well as
the value for $\phi_{S,0}$. The question we would like
to ask is the following: Can the lower bound on $v_S$ be
so small as to allow $v_S$ to be as little as O(eV)? The
arguments presented below suggest that this might not be the
case.

%First, the parameters $a,b$ are chosen phenomenologically
%as to make the potential $V(\phi_S)$ very flat. As such,
%there is of course a great deal of arbitrariness in the
%choices. For the purpose of illustration, $V(\phi_S)$
%shown in Fig. 1 uses $a= 3.37$ and $b = 3.3$, keeping in 
%mind that many other choices are possible. 
To estimate what $\phi_{S,0}$ might be, we now return to 
the issue of the sterile 
neutrino. Let us
assume that there is a Yukawa coupling between the sterile
neutrino, denoted by $\nu_S$, and $\phi_S$, of the form:
\begin{equation}
{\cal L}_{S} = g_{S} \bar{\nu}_{SL} \phi_{S} \nu_{RS} + h.c. .
\end{equation}
When the phase transition has been completed, the sterile
neutrino (Dirac) mass would be $m_{\nu_S} = g_{S} v_S$. But
while $\phi_S$ is ``coasting'' towards its global minimum,
the effective Dirac mass of the sterile neutrino would be given by
\begin{equation}
m_{\nu_S, eff} = g_{S} \phi_{S,0},
\end{equation}
where $\phi_{S,0}$ is the present value of the {\em classical} 
field. As alluded to in the Introduction, there are several
reasons to think that there might be a sterile neutrino. 
However, its mass will depend on a particular scenario.
For definiteness, we shall assume that $m_{\nu_S, eff} \sim
10^{-3} eV$, keeping in mind that other values which are less
than 1 eV are possible.
%For example, if the LSND results hold up in the next few
%years, the most attractive scenario would be one in which
%the sterile neutrino is almost degenerate in mass with the
%electron neutrino, namely $m_{\nu_S, eff} \sim m_{\nu_e}
%$\sim 10^{-3} eV$. 
Also, for the sake of argument, let us
assume that $g_{S} \sim O(1)$ and thus
$\phi_{S,0} \sim O(10^{-3} eV)$. This would then 
imply that
\begin{equation}
\label{phio}
\sqrt{\phi_{S,0} m_{Pl}} \sim 3\: TeV. 
\end{equation}

Next, one might want to see if there is any reason
for the lower bound on $v_S$ to be as low as O(eV).
For this to happen and taking into account\ 
(\ref{phio}, \ref{vS}), one would need $a-b \sim 10^{-48}$,
which means that they are either degenerate to 48 decimal 
places, or that they are as small as $10^{-48}$. This is 
unlikely and undesirable for the following reasons.
First, although the potential
is purely phenomenological, there is no reason
to expect $a$ and $b$ to be of that nature. In fact, 
as we have discussed
above, for a given set of $a,c,d$, the conditions
$V(\phi_S = v_S)= 0$, $V^{\prime}(\phi_S = v_S)=0$
constrain $b$ to be $b = (d/2 -1) exp(c) /c$. Hence,
there is no reason to expect $b$ and $a$ to be degenerate
to 48 decimal places. It is also highly unnatural to expect
both $a$ and $b$ to be of O($10^{-48}$). We shall henceforth
assume that both $a$ and $b$ are of O(1), with $b$
obeying the minimum constraint discussed above.
With $a=3.37, b=3.3$ ($c= 4.5, d=2.33333$) chosen for 
the sole purpose of illustration,
one obtains
\begin{equation}
\label{bound}
v_S \gtrsim 1\: TeV.
\end{equation}
This bound opens up a whole host of interesting possibilities.

First, one could not help but notice an interesting point
which might possibly have a deeper 
meaning. For the sake of argument, let us simply assume
that $v_S \sim 3 TeV$, for example. Then, according to the above 
analysis, this means that we are in a midst of an era where
the scalar field $\phi_S$ is ``slowly rolling'' toward
its global minimum value. The present universe is dominated
by the vacuum energy $V(0) = \sigma^4 \sim (10^{-3} eV)^4$,
implying $\sigma \ll v_S$. It then appears that $\sigma$ 
and $v_S$ are completely
unrelated (which might still be the case in a deeper theory).
However, one notices that 
\begin{equation}
\label{sigma}
v_S^2 / m_{Pl} \sim 10^{-3} eV.
\end{equation}
if $v_S \sim 3 TeV$. Does this numerical exercise
imply that $\sigma \equiv v_S^2 / m_{Pl}$? After all,
in our scenario, there are three scales: $\sim 10^{-3} eV$,
$\sim 1 TeV$, and $m_{Pl}$. It might not be too surprising
that one of the scales (e.g. $\sim 10^{-3} eV$) is related
to the other two.
This intriguing possibility prompts us to rewrite the
phenomenological potential $V(\phi_S)$ as
\begin{equation}
V(\phi_S) = (v_S^2 / m_{Pl})^4 (1- a x^2 + 
b x^2 exp(-c x^2) + d x^3),
\label{potential2}
\end{equation}
where $b = (d/2 -1) exp(c) /c$ and where $v_S$ is of order of
a few TeV's. The value of $v_S$ which is determined from
the constraint of ``slow rolling'' and the presumed
mass of a sterile neutrino, independently of the
value of the vacuum energy, can in turns be used to {\em fix} 
the vacuum energy itself if the potential has the form\ 
(\ref{potential2}). What might be the origin of such a potential
and of such a scale? Is that scale related to the scale
of extra dimensions or of SUSY breaking?
These questions are beyond the scope of this paper.

To complete the discussion, let us make a rough estimate
of the time it takes to get from the present era to
the point $\phi_{S,e}$ where $\phi_S$ starts to evolve rapidly. 
To be specific,let us take $\phi_{S,0} \sim O(10^{-3} eV)$ and
$v_S \sim 3 \, TeV$. From\ (\ref{bound}), $\phi_{S,e}$ will
be approximately $3\, 10^{-3} eV$. This means that we will be
looking at the evolution from $x_0 \sim 10^{-31}$ to
$x_e \sim 10^{-30}$, where $x \equiv (\phi_S/v_S)^2$. Let us
also make the approximation that $H$ is constant
($\approx H_0$) during that
period (it does not vary much as we have shown above). One can
then estimate the additional time $\Delta t \equiv t_e - t_0$
($t_0$ is the present time) 
where the slow rolling begins to
be invalid. It is straightforwardly given by :
\begin{equation}
\label{time}
 \Delta t \approx  \frac{2\pi}{H_0}\frac{v_S^2}{m_{Pl}^2} 
\int_{x_0}^{x_e} \frac{\tilde{V}}{-x \tilde{V}^{\prime}} dx, 
\end{equation}
where $\tilde{V}$ has been defined as $V = \sigma^4 \tilde{V}$,
and where the second on the right-hand side of Eq.\ (\ref{time})
is a rewriting of the well-known formula used to compute 
the number of e-folds in an inflationary scenario. A numerical
integration of Eq.\ (\ref{time}) gives
$\Delta t \equiv t_e - t_0 \approx 36 /H_0$. With $1/H_0 \sim
15 \times 10^9$ yr, one obtains roughly $\Delta t \sim 540 \times
10^9$ yr. The universe will still be accelerating for a long, 
long time! (It turns out that $\Delta t$ gets even bigger as
$v_S$ gets larger.) In fact, the universe will undergo an
inflationary period until the phase transition is completed.
The small latent heat will be converted into the production
of very massive (TeV) $\phi_S$, which could decay into very massive
$\nu_S$ if the masses allow for it to be so.
%Of course, this fate of the universe is
%probably an academic exercise. Nevertheless, it is still
%a prediction of this particular scenario.

%What happens when the phase transition is about to be completed?
%The small latent heat will be converted into the production
%of very massive (TeV) $\phi_S$, which could decay into very massive
%$\nu_S$ if the masses allow for it to be so. It is not clear what
%implications this could have on the universe at that time except
%the fact that $\nu_S$ is a singlet, with very little mixing
%with one of the active neutrinos ($\nu_e$).

A couple of other remarks are in order here.
Since $\sqrt{V^{\prime \prime}(\phi_{S,0}})$ is the present
effective mass of the Higgs field $\phi_S$, the slow rolling
condition would indicate that this effective mass would be
at most $10^{-33} eV$. Would such a low mass cause any
problem with long range forces? There are strong 
constraints on the couplings of
such a low mass scalar to ordinary matter \cite{quint}. 
In our case, this
singlet scalar field only couples to another singlet neutrino,
$\nu_S$, which in turns has a tiny mixing with $\nu_e$. The
effects of this low mass scalar on ordinary matter appear
to be negligible. A rough estimate of the coupling of
$\phi_S$ to an electron via a W-mediated loop diagram gives
an efective coupling $\sim G_{F} m_{\nu_S}^2 |V_{e\nu_S}|^2$.
With $|V_{e\nu_S}|^2 \sim 10^{-3}$ and $m_{\nu_S}^2 \sim
10^{-6} eV^2$, one expects this coupling to be less than 
$10^{-30}$. When it is squared to provide the coupling for
a ``long range'' potential, one expects it to be $\sim
10^{-60}$, and this is considerably less than $G m_{e}^2/
m_{Pl}^2 \sim 10^{-44}$ for the gravitational potential.
For all practical purposes, the effective coupling to
matter (electron) is so weak that one can safely ignore it.
This subject will be dealt with in future work. Another
topic of interest for a future work is the effect of such light
scalars in supernovae process (such as the r-process for example).
A preliminary investigation indicates that, because of
the extreme weakness of the interaction, there is practically
no effect. 
The second remark has to do with the question of why $\Omega_m
\sim 0.3$ and $\Omega_{\Lambda} \sim 0.7$ differ only by a
factor of 2 or so at the present time. Tracker models of
quintessence \cite{quint}are supposed to 
address this issue, although
it is still controversial about the magnitude of 
$\rho_{vacuum}$ (fine-tuning problem). At this point,
the statement that one can make about the scenario presented
here is the following: Some yet-unknown physics gives rise
to a potential of the form\ (\ref{potential2}) with a scale
$v_S$ of a few TeV's, so that if $\Omega_{\Lambda} \sim 0.7$
then $\Omega_{m}(t_0) + \Omega_{\Lambda} (t_0) =1$ implies
$\Omega_m \sim 0.3$. These issues are certainly beyond the
scope of this paper.

In conclusion, we have presented a scenario in which the physics
of the {\em accelerating universe} is intrincately linked to that of
{\em a sterile neutrino}.  
It is amusing to note that, although
the idea of inflation appears to be strengthened by the new
astrophysical results, it must have happened at the dawn of
the universe. The fact that the present universe appears to
be accelerating prompts us to think that we are starting
to {\em experience a late-time inflation}, of a different
nature from the one at the birth of the present universe.  
In addition, the dark energy density of our picture is
truly a constant, $V(\phi_S=0)$, in contrast with scenarios
based on quintessence in which it is time-dependent. This is 
something which could be tested within (hopefully) the 
next ten years.

I would like to thank Marc Sher for bringing my attention to
an earlier paper \cite{singh} whose motivation was similar 
to the one presented here: the link of the physics of 
neutrinos to the present dark energy, and for useful comments. 
That model is however
completely different from ours. To the best of our knowledge,
the model presented here is the first attempt 
to link the dark energy to the physics
which gives rise to the mass of the 
{\em sterile neutrino}. My thanks
also go to Paul Frampton  and Manfred Lindner for useful comments.
 
This work is supported in parts by the US Department
of Energy under grant No. DE-A505-89ER40518.

\end{document}